\documentclass[5p, number]{elsarticle}

\usepackage{graphicx}
\usepackage{amssymb}
\usepackage{amsmath}

\biboptions{sort&compress}
\begin{document}
\title{UV spectra of iron-doped carbon clusters FeC$_n$ $n$ = 3--6 \\ \vspace{0.2cm} \small{\textit{published in} International Journal of Mass Spectrometry 365--366 (2014) 351--355}}

\author{Mathias Steglich} \ead{m.steglich@web.de}
\author{Xiaojing Chen}
\author{Anatoly Johnson}
\author{John P. Maier} \ead{j.p.maier@unibas.ch}
\address{Department of Chemistry, University of Basel, Klingelbergstrasse 80, 4056 Basel, Switzerland}

\date{\today}

\begin{abstract}
Electronic transitions of jet-cooled FeC$_n$ clusters ($n = 3 - 6$) were measured between 230 and 300 nm by a mass-resolved 1+1 resonant two-photon ionization technique. Rotational profiles were simulated based on previous calculations of ground state geometries and compared to experimental observations. Reasonable agreement is found for the planar fan-like structure of FeC$_3$. The FeC$_4$ data indicate a shorter distance between the Fe atom and the bent C$_4$ unit of the fan. The transitions are suggested to be $^{3}$A$_{2} \leftarrow ^{3}$B$_{1}$ for FeC$_3$ and $^{5}$A$_{1} \leftarrow ^{5}$A$_{1}$ for FeC$_4$. In contrast to the predicted C$_{\infty \text{v}}$ geometry, non-linear FeC$_5$ is apparently observed. Line width broadening prevents analysis of the FeC$_6$ spectrum.
\end{abstract}

\maketitle

\section{Introduction}
The structures and electronic properties of carbonaceous molecules and nanoparticles doped with transition metals have received increasing interest in the past years due to potential applications in nanotechnology. Early 3$d$ and 4$d$ transition metals (M $=$ Ti, Zr, V, Nb, Cr, Mo) form metallo-carbohedrenes (met-cars), very stable clusters with the molecular formula M$_8$C$_{12}$ \cite{castleman92, wei92, wei92b, pilgrim93, deng94}. Late transition metals (Fe, Co, Ni) act as catalysts in carbon nanotube growth \cite{kumar10}, but only iron does also form met-cars, whereby several stable structures have been detected by mass spectrometry, in particular Fe$_7$C$_8$, Fe$_8$C$_{12}$, Fe$_{12}$C$_{12}$ \cite{pilgrim93}.

Iron-carbon compounds might also play an important role in astrochemistry as Fe and C belong to the most abundant refractory elements in the Galaxy. In addition, iron is heavily depleted in the interstellar medium (ISM) \cite{olthof75, deboer78, savage79}. It must be trapped in dust grains or in molecules. So far, FeCN and possibly FeO are the only iron-containing molecules that have been detected in the ISM \cite{zack11, walmsley02}. In this context, more laboratory spectra, especially for larger species and at optical wavelengths, are needed.

Large iron-carbon clusters ($400 - 1000$ amu) were investigated by mass spectrometry \cite{pilgrim93, huisken00} and density functional theory (DFT) \cite{cune01, harris07, gutsev12, ryzhkov12}. Smaller molecules ($< 153$ amu) were studied experimentally by mass spectrometry \cite{lebrilla87, drewello90}, infrared matrix isolation spectroscopy \cite{kafafi85, kline85, ball93, chu00}, anion photoelectron spectroscopy \cite{fan94, drechsler95, fan95, li99, wang00, drechsler03}, and gas phase ion chromatography \cite{helden94}. High-resolution spectra at optical and mm wavelengths were measured only for FeC \cite{balfour95, allen96, brugh97, fujitake01, aiuchi01, steimle02}. Several small Fe$_n$C$_m$ ($n < 8$, $m < 11$) species have been the subject of theoretical calculations \cite{cao96, nash96, arbuznikov99, tzeli02, gutsev03, noya03, hendrickx04, ryzhkov05, rayon06, yuan06, ma07, ryzhkov08, sun08, largo09, redondo09, zhu09}. The structures and electronic properties of the small clusters are important to understand the growth mechanism of larger compounds, such as met-cars and nanotubes. However, the prediction of the correct ground state geometries is difficult due to the large number of local minima on the potential energy surfaces.

\begin{figure}\begin{center}
 \includegraphics[scale=0.1]{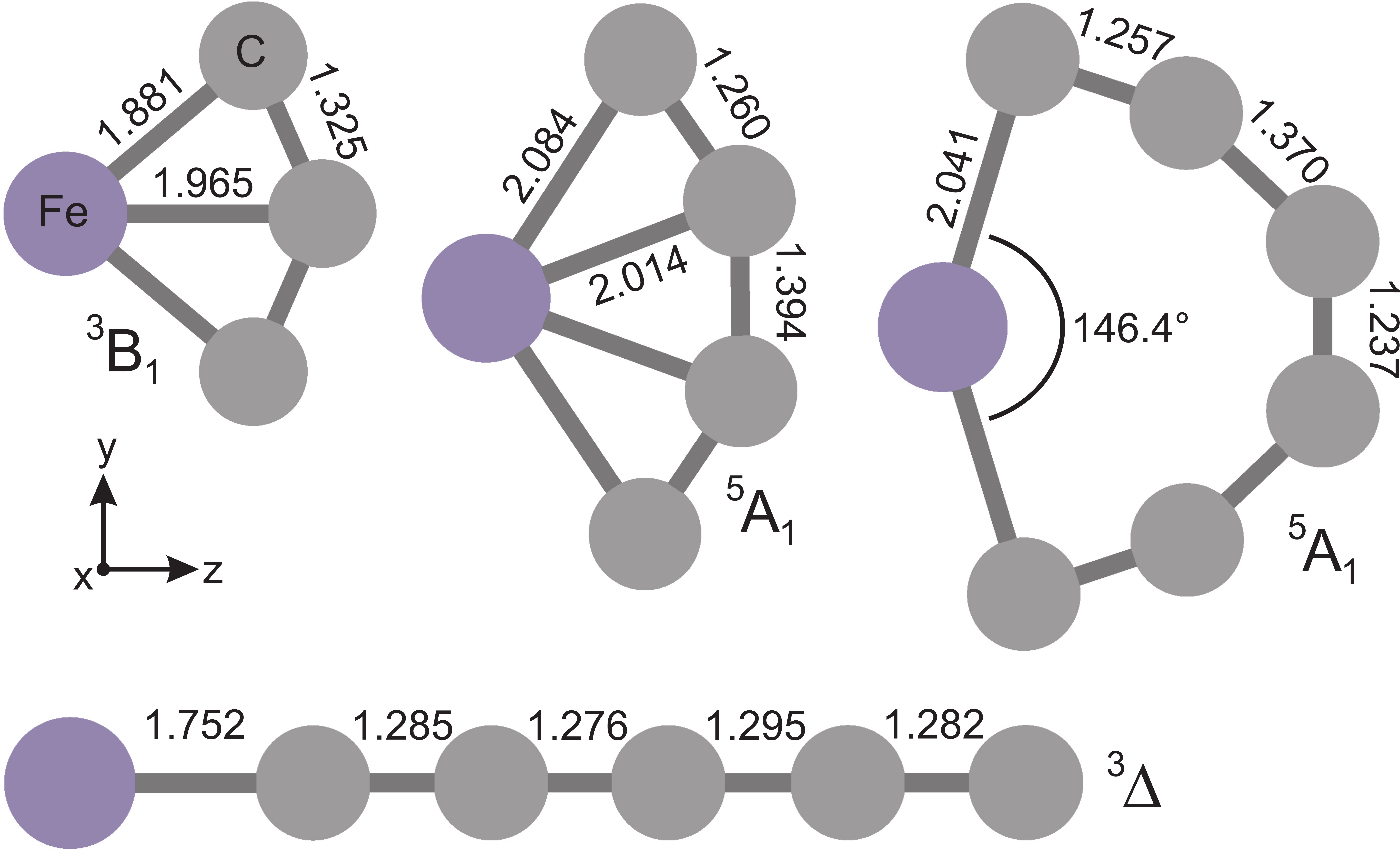}
 \caption{Calculated ground state structures of FeC$_n$ \cite{largo09,zhu09}, with bond lengths in \AA. \label{fig_structures}}
\end{center}\end{figure}

In this article, gas phase spectra of jet-cooled FeC$_n$ clusters ($n = 3 - 6$) between 230 and 300\,nm are presented. Conclusions concerning the molecular structures are drawn. The most recent calculations of the ground state geometries of these molecules applied the DFT functional B3LYP in conjunction with the 6-311+G($d$) basis set \cite{largo09,zhu09}. It was found that the minimum energy structure is planar fan-like (C$_{2\text{v}}$) for FeC$_3$ and FeC$_4$, linear (C$_{\infty\text{v}}$) for FeC$_5$, and  cyclic (C$_{2\text{v}}$) for FeC$_6$ (Fig.\,\ref{fig_structures}). In all these cases, the C$_n$ unit is not disrupted as C--C bonds are much stronger than Fe--C ones. The only experimental data available so far are photoelectron spectra of the FeC$_3$ and FeC$_4$ anions, whereby contradictory information regarding the geometries of the neutrals (C$_{\infty\text{v}}$ vs. C$_{2\text{v}}$) was obtained \cite{fan95, wang00}.

\section{Experimental}
The setup used a molecular beam source combined with a linear time-of-flight (TOF) mass spectrometer (MS). The source consists of a pulsed valve coupled with laser vaporization. Ablation of a rotating and translating iron rod was achieved by focusing 532\,nm radiation (20\,mJ, 5\,ns) to $\approx$\,300\,$\mu$m. The iron vapor was co-expanded in a He/2\,\%\,C$_2$H$_2$ gas mixture (8\,bar backing pressure) and the supersonic jet was skimmed 40\,mm downstream to produce a collimated beam. A +300 V potential was applied to the skimmer to remove ions before neutrals entered the ionization region of the TOF-MS. Rotational temperatures of the FeC$_n$ species in the molecular beam were below 50\,K (Sect. \ref{sec_results}); translational temperatures are expected to be $< 1$\,K.  Vibrational cooling, however, is not very efficient in supersonic jet expansions. The corresponding temperatures can exceed 500\,K.

\begin{figure}\begin{center}
 \includegraphics[scale=0.44]{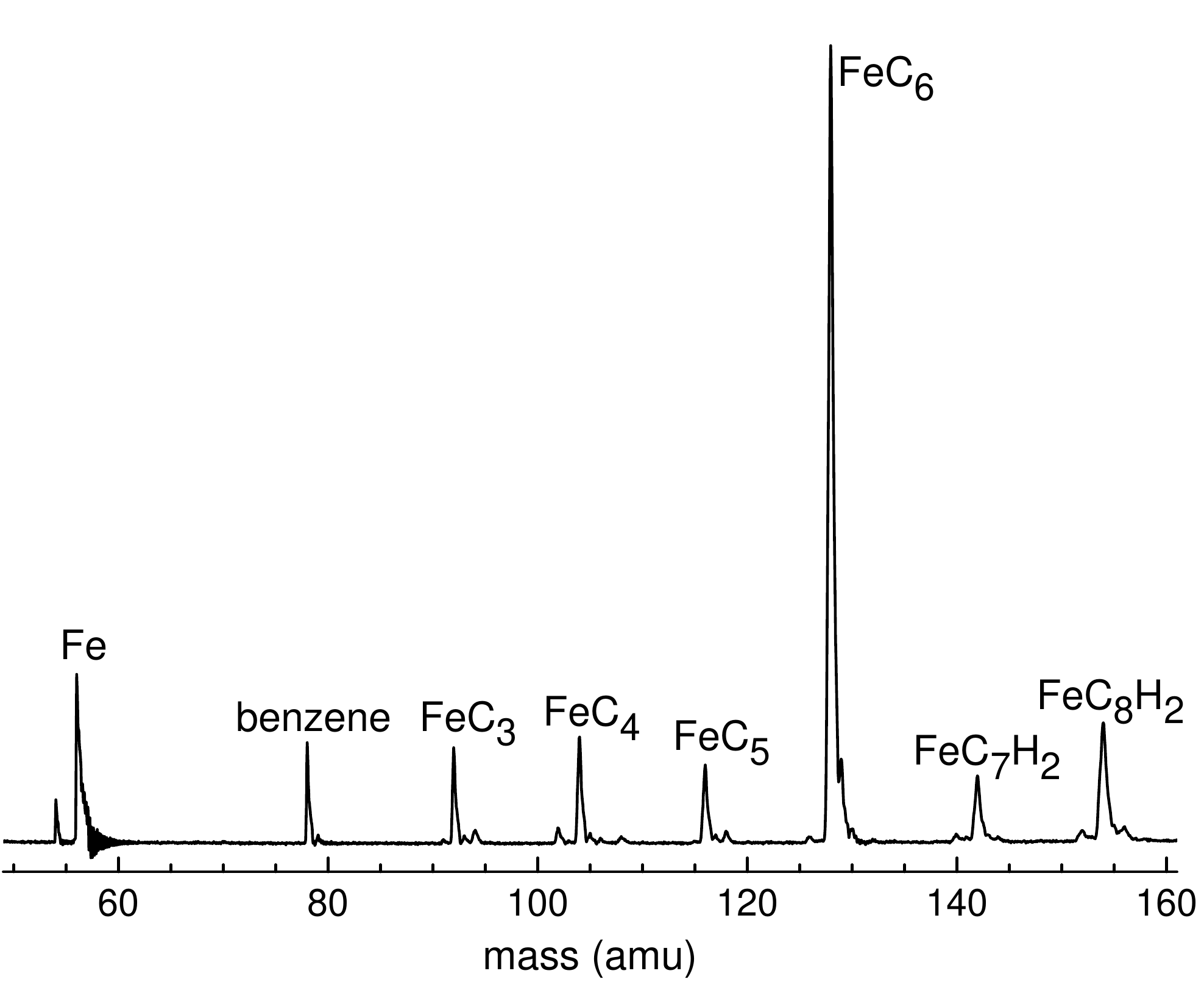}
 \caption{Integrated mass spectrum observed during the 1+1 R2PI scan ($267.62 - 267.27$\,nm). \label{fig_TOF}}
\end{center}\end{figure}

Spectral scans applying a 1+1 resonant two-photon ionization (R2PI) scheme were realized by coun\-ter-pro\-pa\-ga\-ting the laser radiation into the molecular beam. Two different laser sources were used. Broad range scans between 230 and 400\,nm were conducted with an optical parametric laser ($5 - 10$\,ns, 20\,Hz, 0.1\,nm bandwidth). The frequency-doubled output of a higher-resolution dye laser ($\sim 10$\,ns, 10\,Hz, 0.001\,nm bandwidth) was used for rotational profile scans. The wavelengths were calibrated with observed lines of atomic iron (0.01\,nm uncertainty). Rotational profiles were fitted with the PGopher software \cite{western10}. The calculated ionization energies of FeC$_n$ ($n = 3 - 6$) are $7 - 9$\,eV \cite{zhu09}. 1+1 R2PI scans are therefore possible for wavelengths below $354 - 275$\,nm. The low-resolution spectra were power-corrected. Negative absorption lines can be noticed at 272 and between 244 and 256 nm. They are caused by atomic iron, by far the most abundant species in the molecular beam, absorbing most of the laser power.

\section{Results and Discussion} \label{sec_results}
An integrated TOF mass spectrum as observed during the higher-resolution scan between 267.62 and 267.27\,nm is displayed in Fig.\,\ref{fig_TOF}. It can be seen that the cluster source produces Fe, benzene, FeC$_n$ ($n = 3 - 6$), and FeC$_n$H$_2$ ($n = 7, 8$). The relative mass peak intensities are not only determined by the species' abundance, but also by the integrated R2PI signal, hence the anomalously high intensity of the FeC$_6$ peak.

\begin{figure*}\begin{center}
 \includegraphics[scale=0.72]{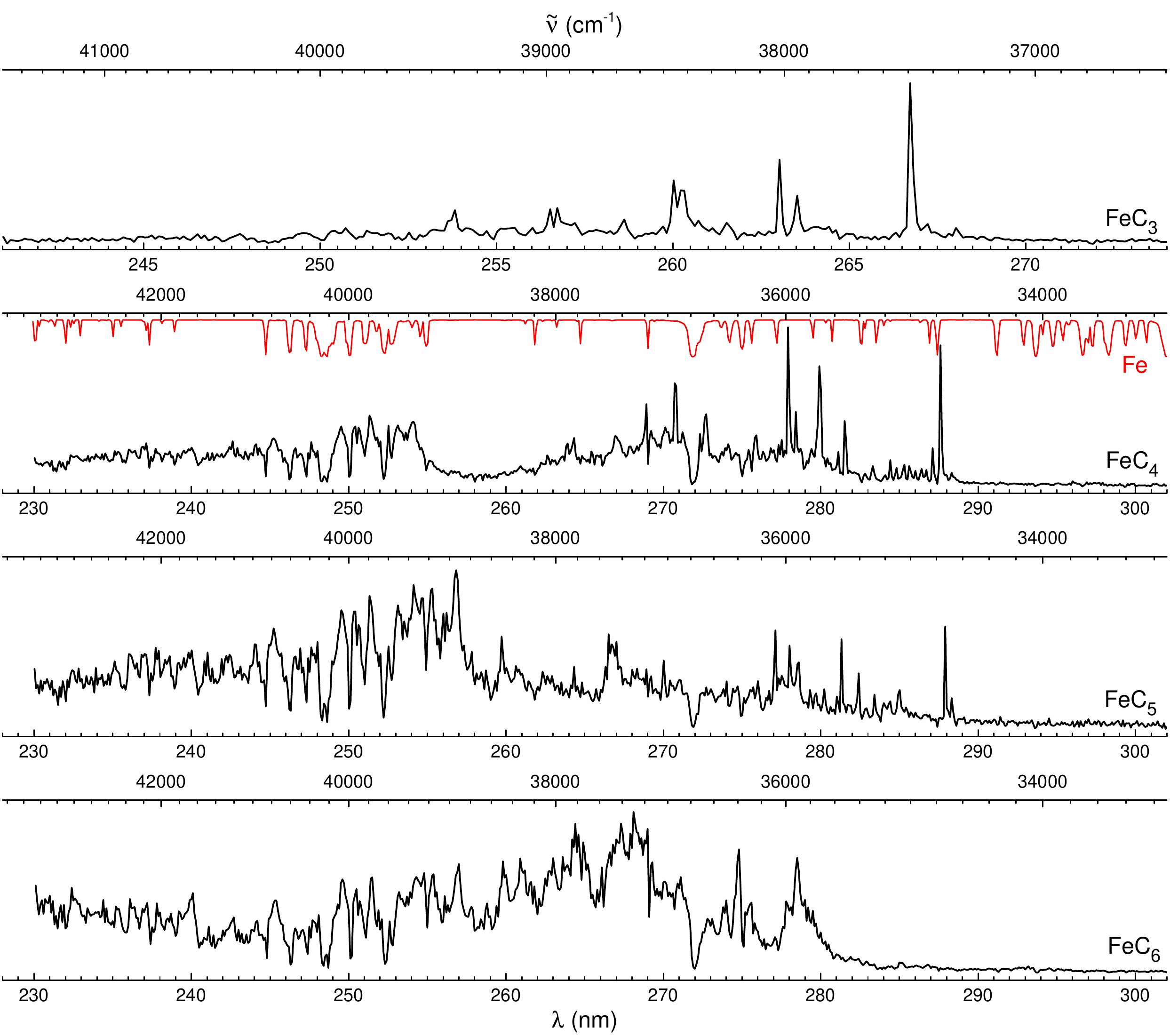}
 \caption{Electronic spectra of FeC$_n$ ($n = 3 - 6$) measured by a 1+1 R2PI method using a low-resolution (0.1\,nm) laser. The spectrum of atomic iron is shown in red. \label{fig_FeCnLR}}
\end{center}\end{figure*}

\begin{figure}\begin{center}
 \includegraphics[scale=0.7]{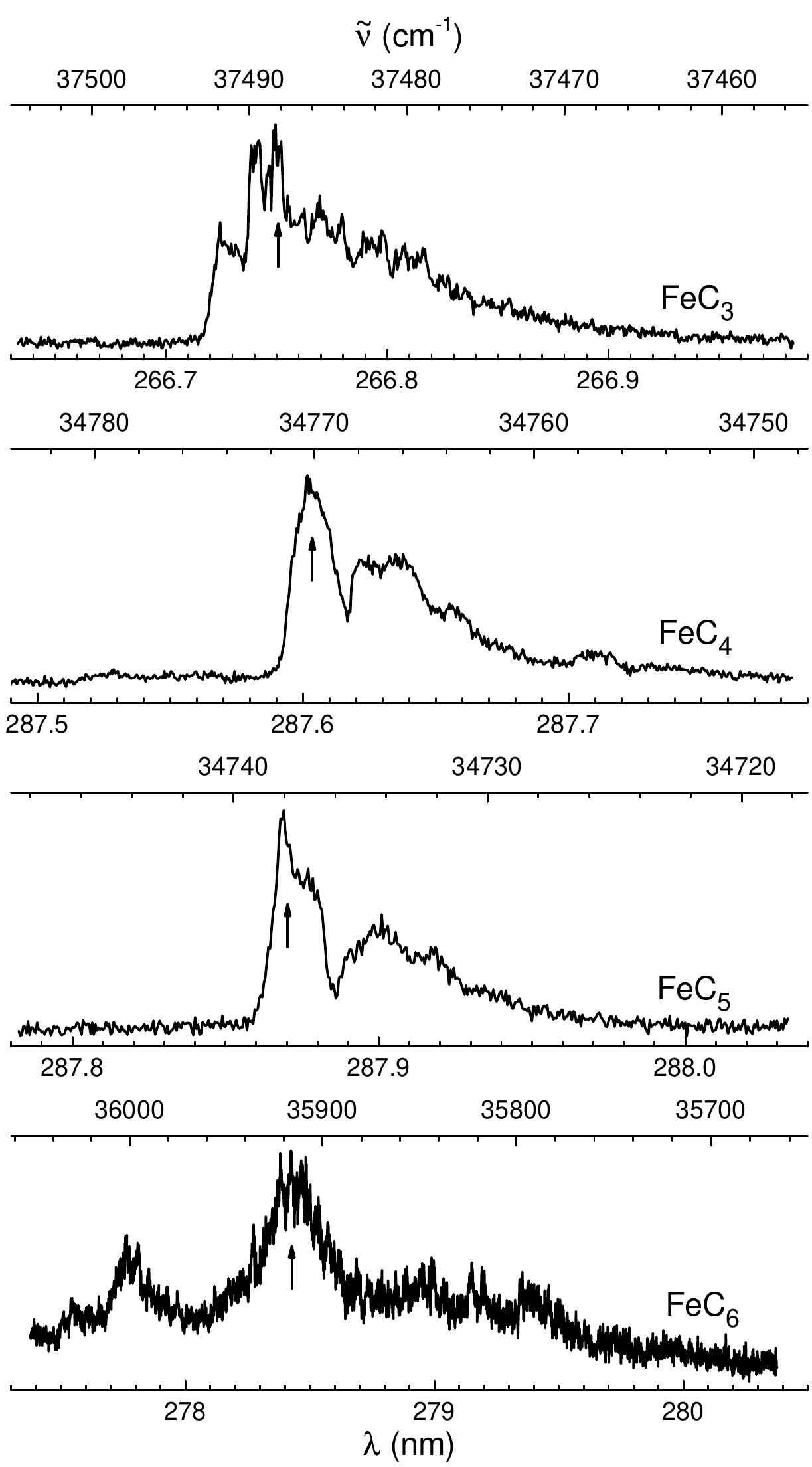}
 \caption{Higher-resolution (0.001\,nm) 1+1 R2PI spectra of the first stronger FeC$_n$ ($n = 3 - 6$) bands. The arrows indicate the absorption maxima summarized in Table\,\ref{tab_1}. \label{fig_FeCnHR}}
\end{center}\end{figure}

Figures \ref{fig_FeCnLR} and \ref{fig_FeCnHR} present the measured electronic spectra of the FeC$_n$ species. The wavelengths of the stronger bands (their maxima) as indicated by arrows in Fig.\,\ref{fig_FeCnHR} are given in Table\,\ref{tab_1}.

\begin{table}\begin{center}
 \caption{Positions of the absorption band maxima in the electronic transitions of the FeC$_n$ species as indicated by arrows in Fig.\,\ref{fig_FeCnHR}.}
\begin{tabular}{lcc} \hline
  FeC$_n$ & $\lambda$ (nm) & $\tilde{\nu}$ (cm$^{-1}$) \\ \hline \hline
  $n = 3$ & 266.75 & 37477 \\
  $n = 4$ & 287.61 & 34760 \\
  $n = 5$ & 287.87 & 34728 \\
  $n = 6$ & 278.43 & 35906 \\ \hline
\end{tabular}\\
\label{tab_1}
 \end{center}\end{table}

\subsection{FeC$_3$}
The strongest band is located at 266.75\,nm and appears to be the origin band of the transition due to the Franck-Condon intensity distribution. Several bands at higher energy correspond to the excitation of a vibrational mode ($470 - 530$\,cm$^{-1}$) in the upper electronic state.

Density functional theory (DFT) and time-dependent DFT (TDDFT), implemented in the Gaussian09 software package~\cite{frisch13}, were applied to predict the vibrations in the ground state and the excited electronic state energies of the previously optimized C$_{2\text{v}}$ structure at the B3LYP/6-311+G($d$) level of theory. A vibrational mode in the ground state of a$_1$ symmetry was calculated at 444\,cm$^{-1}$ (unscaled value), which agrees reasonably well with the observed vibrational progression in the excited state. The TDDFT computations predict two electronic transitions with notable oscillator strength in the observed wavelength region. A $\text{T}_{31}(^{3}\text{A}_{2}) \leftarrow \text{T}_0(^{3}\text{B}_{1}$) transition ($f = 0.027$) was calculated at 263.3\,nm and a $\text{T}_{32}(^{3}\text{A}_{1}) \leftarrow \text{T}_0(^{3}\text{B}_{1}$) transition ($f = 0.034$) at 260.6\,nm. Further transitions are to be expected at longer wavelengths. However, the applied one-colour technique allows the detection of electronic transitions only down to 273.7\,nm because the calculated vertical ionization energy is 9.06\,eV. One finds similar results with calculations using the larger basis set aug-cc-pvqz: $\text{T}_{31}(^{3}\text{A}_{2}) \leftarrow \text{T}_0(^{3}\text{B}_{1}$) at 266.5\,nm with $f = 0.026$ and $\text{T}_{32}(^{3}\text{A}_{1}) \leftarrow \text{T}_0(^{3}\text{B}_{1}$) at 261.5\,nm with $f = 0.028$.

\begin{figure}\begin{center}
 \includegraphics[scale=0.44]{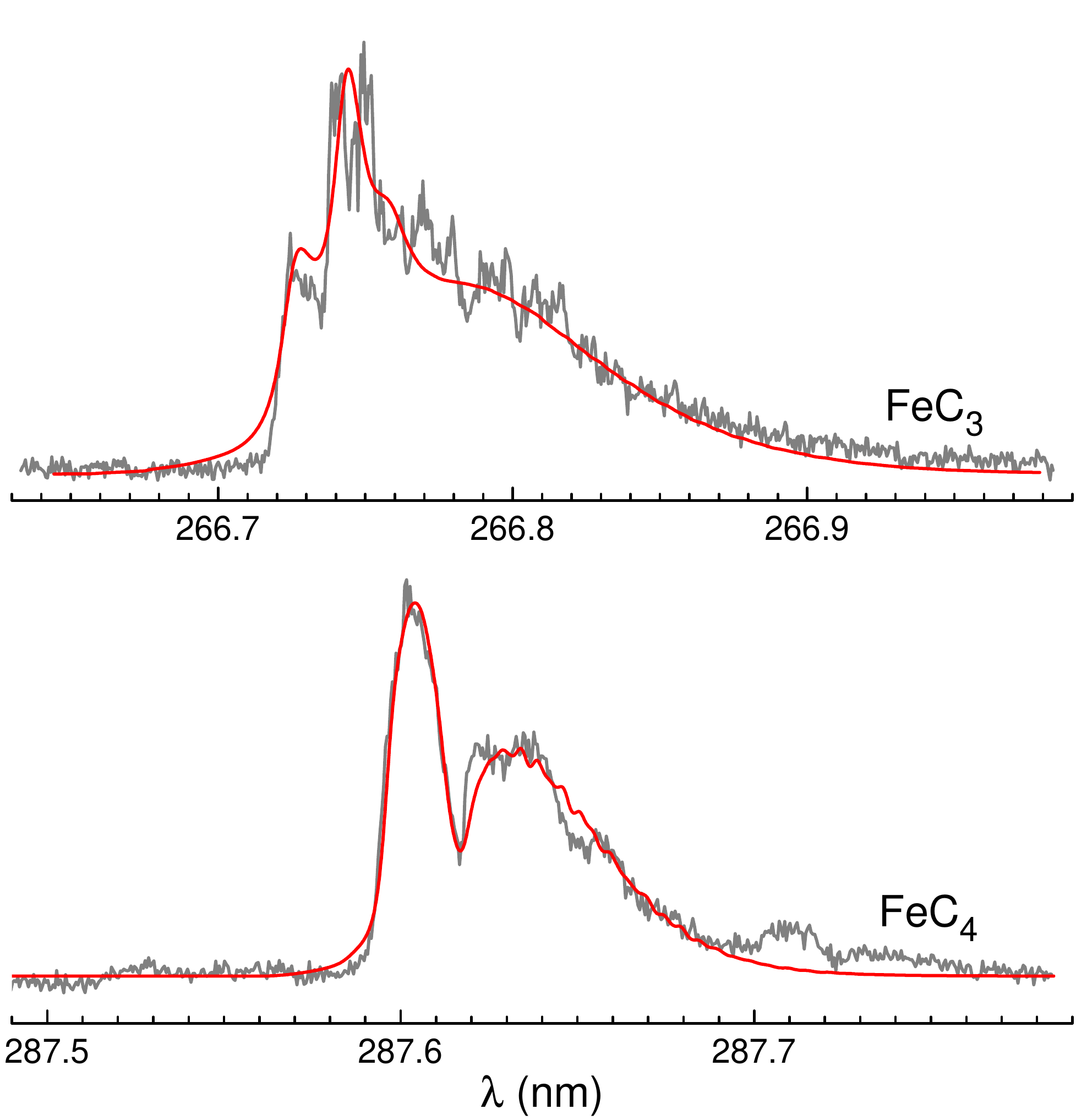}
 \caption{Experimentally measured (grey) and simulated (red) rotational profiles of FeC$_3$ ($\text{A}_{2} \leftarrow \text{B}_{1}$) and FeC$_4$ ($\text{A}_{1} \leftarrow \text{A}_{1}$) bands. \label{fig_fits}}
\end{center}\end{figure}

The theoretical ground state geometry corresponds to rotational constants $A = 0.476$\,cm$^{-1}$, $B = 0.285$\,cm$^{-1}$, and $C = 0.178$\,cm$^{-1}$. The rotational profile of the absorption band at 266.75\,nm can be simulated by a C$_{2\text{v}}$ asymmetric top molecule. The best fit to the observed profile is achieved for a $\text{A}_{2} \leftarrow \text{B}_{1}$ transition (Fig.\,\ref{fig_fits}). The noise in the high-intensity part of the experimental profile is caused by instabilities of the laser vaporization source. A rotational temperature of 45(15)\,K is indicated from the fit and the ground state rotational constants are 0.479(0.010)\,cm$^{-1}$, 0.286(0.010)\,cm$^{-1}$, and 0.179(0.010)\,cm$^{-1}$, in good agreement with theory. The rotational constants in the excited state are lower by less than 10\,\%. In view of these results, the observed band system is suggested to be the $^{3}$A$_{2} \leftarrow ^{3}$B$_{1}$ transition of FeC$_3$ with planar fan-like structure shown in Fig.\,\ref{fig_structures}.

\subsection{FeC$_4$}
The first strong band in the FeC$_4$ electronic spectrum appears at 287.61\,nm. TDDFT calculations at the B3LYP/6-311+G($d$) level of theory predict a $^{5}$A$_{1} \leftarrow ^{5}$A$_{1}$ transition at 287.0\,nm ($f = 0.038$). Additional transitions with reasonable oscillator strengths are calculated at 262.9\,nm ($^{5}$A$_{1} \leftarrow ^{5}$A$_{1}$; $f = 0.017$), 261.5\,nm ($^{5}$B$_{1} \leftarrow ^{5}$A$_{1}$; $f = 0.014$), 256.0\,nm ($^{5}$A$_{1} \leftarrow ^{5}$A$_{1}$; $f = 0.009$), and 229.7\,nm ($^{5}$B$_{2} \leftarrow ^{5}$A$_{1}$; $f = 0.080$), as well as at wavelengths longer than 300\,nm inaccessible with the applied 1+1 R2PI used in this work.

The rotational constants of the calculated ground state structure are $A = 0.253$\,cm$^{-1}$, $B = 0.202$\,cm$^{-1}$, and $C = 0.113$\,cm$^{-1}$. Profile fits, assuming an asymmetric top, can be obtained for any of the three possible transition types, $\text{A}_{1} \leftarrow \text{A}_{1}$, $\text{B}_{1} \leftarrow \text{A}_{1}$, and $\text{B}_{2} \leftarrow \text{A}_{1}$. The best fit is achieved for an $\text{A}_{1} \leftarrow \text{A}_{1}$ transition (Fig.\,\ref{fig_fits}). The inferred parameters for the different transition types are listed in Table\,\ref{tab_2}. In all the three cases, the $C$ constant in the ground state is bigger than the theoretical value. This suggests a shorter distance between the Fe atom and the center of the bent C$_4$ unit than predicted by B3LYP/6-311+G($d$).

\begin{table*}\begin{center}
 \caption{Rotational constants and temperature inferred from the profile of the 287.61\,nm absorption band in the  R2PI spectrum of FeC$_4$.}
\begin{tabular}{lcccc} \hline
transition & $A$ (cm$^{-1}$) & $B$ (cm$^{-1}$) & $C$ (cm$^{-1}$) & $T_{\text{rot}}$ (K) \\ \hline \hline
$\text{A}_{1} \leftarrow \text{A}_{1}$ & 0.250$\leftarrow$0.298 & 0.167$\leftarrow$0.176 & 0.160$\leftarrow$0.170 & 18 \\
$\text{B}_{1} \leftarrow \text{A}_{1}$ & 0.208$\leftarrow$0.246 & 0.177$\leftarrow$0.187 & 0.201$\leftarrow$0.216 & 23 \\
$\text{B}_{2} \leftarrow \text{A}_{1}$ & 0.204$\leftarrow$0.243 & 0.223$\leftarrow$0.241 & 0.212$\leftarrow$0.229 & 24 \\ \hline
\end{tabular}\\
\label{tab_2}
 \end{center}\end{table*}

The observed band at 287.61\,nm is thus tentatively assigned as the origin band of a $^{5}$A$_{1} \leftarrow ^{5}$A$_{1}$ transition. The other bands at shorter wavelengths can be vibrational excitation in the upper state or belong to another electronic transition.

\subsection{FeC$_5$}
In contrast to FeC$_3$ and FeC$_4$, a linear ground state geometry was calculated for FeC$_5$ \cite{largo09,zhu09}, which yields a rotational constant of 0.023\,cm$^{-1}$.  Using this constant for the ground state and a variable one for the excited state, it was not possible to simulate the entire rotational profile of the band at 287.87\,nm. A fit could perhaps be obtained by using the right combination of constants for spin-orbit coupling, spin-spin coupling, and $\Lambda$ doubling. However, in comparison with the observed absorption profile of FeC$_4$ at 287.61\,nm, it seems more likely that a non-linear molecule is observed. It should be noted that three non-linear geometries with C$_{2\text{v}}$ or close to C$_{2\text{v}}$ symmetry and with different electronic ground states ($^5$B$_1$, $^3$A', and $^5$A$_1$) are predicted by theory to be within 40\,kJ/mol of the linear structure \cite{largo09,zhu09}.

\subsection{FeC$_6$}
The first strong feature in the measured spectrum of FeC$_6$ appears at 278.43\,nm. It is very broad ($\approx 37$\,cm$^{-1}$) and has an almost perfect Lorentzian shape. Assuming lifetime broadening to be responsible for the observed width, a lower limit of about 0.1\,ps for the lifetime of the excited electronic state is deduced.

\section{Conclusions}
Iron-containing carbonaceous molecules with FeC$_n$ ($n = 3 - 6$) stoichiometry, which have been predicted only by theory so far, were produced under laboratory conditions. Their electronic spectra were measured between 230 and 300\,nm by a mass-resolved 1+1 resonant two-photon ionization technique. Higher-resolution scans of the first strong bands were compared with simulated rotational profiles using ground state structures previously calculated at the B3LYP/6-311+G($d$) level \cite{largo09,zhu09}. Observed and predicted profiles agree reasonably well for a fan structure of FeC$_3$. The transition with origin band at 266.75 nm is assigned as $^{3}$A$_{2} \leftarrow ^{3}$B$_{1}$. The distance between the bent carbon chain and the Fe atom in the FeC$_4$ fan seems to be shorter than calculated. A band measured at 287.61 nm is probably the origin band of a $^{5}$A$_{1} \leftarrow ^{5}$A$_{1}$ transition. The measured profile of FeC$_5$ indicates that a non-linear molecule was observed, in contrast to the predicted linear geometry. No conclusions can be drawn from the FeC$_6$ electronic spectrum due to line-width broadening.

The measurements presented here do not prove conclusively certain molecular structures. The possible presence of other isomers has not been evaluated yet due to the complexity of the UV spectra. Additional spectroscopic studies are needed to further explore the structural properties and the low-lying electronic states.

\section*{Acknowledgements}
This work has been funded by the Swiss National Science Foundation (Project 200020-140316/1).

\bibliography{references}

\end{document}